\documentclass[11pt,a4paper]{article}

\usepackage{amsmath}
\usepackage{amsfonts}
\usepackage{amssymb}
\usepackage{graphicx,psfrag,color}
\usepackage{subfigure}
\usepackage{hyperref}

\usepackage{enumerate}

\usepackage[left=2cm,top=2.5cm,right=2cm,bottom=2cm]{geometry}

\begin{document}
\begin{center}
\Large{\textbf{Cosmology with hybrid expansion law:  scalar field reconstruction of cosmic history and observational constraints}} \\[0.5cm]
 
\large{\textbf{\"{O}zg\"{u}r Akarsu$^{\rm a}$,  Suresh Kumar$^{\rm b}$, R. Myrzakulov$^{\rm c}$,   M. Sami$^{\rm c}$,   Lixin Xu$^{\rm d}$}}
\\[0.5cm]

\small{
\textit{$^{\rm a}$ Department of Physics, Ko\c{c} University, 34450 Sar{\i}yer, {\.I}stanbul, Turkey.}}

\vspace{.2cm}

\small{
\textit{$^{\rm b}$ Department of Mathematics, BITS Pilani, Pilani Campus, Rajasthan-333031, India.}}

\vspace{.2cm}

\small{
\textit{$^{\rm b}$ Centre of Theoretical Physics, Jamia Millia Islamia, New Delhi-110025, India.}}

\vspace{.2cm}

\small{
\textit{$^{\rm c}$ Institute of Theoretical Physics, Dalian University of Technology, Dalian, 116024, P. R. China.}}

\end{center}
\textbf{E-Mail:} oakarsu@ku.edu.tr,  sukuyd@gmail.com, rmyrzakulov@gmail.com, samijamia@gmail.com,  lxxu@dlut.edu.cn

\vspace{.8cm}

\noindent \small{\textbf{\large{Abstract.}} 
In this paper, we consider a simple form of
expansion history of Universe referred to as
the hybrid expansion law $-$ a product of power-law
and exponential type of functions. The ansatz by
construction mimics the power-law and de Sitter
 cosmologies as special cases but also provides an
 elegant description of the transition  from deceleration
 to cosmic acceleration. We point out the Brans-Dicke
  realization of the cosmic history under consideration.
  We construct potentials for quintessence, phantom and
  tachyon fields, which can give rise to the hybrid
  expansion law in general relativity.  We investigate  observational constraints
   on the model with hybrid expansion law applied to late time acceleration as well
    as to early Universe {\it a la nucleosynthesis}.\\

\noindent \small{\textbf{\large{Keywords.}} dark energy theory, supernova type Ia - standard candles, big bang nucleosynthesis

%\tableofcontents

%\vspace{0.3cm}

\section{Introduction}
\label{Intro}

Since the first observation on late time cosmic acceleration in 1998 \cite{Riess98,Perlmutter99}, attempts have been made to understand the cause of this remarkable phenomenon within the framework of Einstein general relativity and beyond it. Broadly, the model building undertaken in the literature to capture the essential features of cosmological dynamics can be classified in two categories: Models based on dark energy \cite{vpaddy,review2,review3, review3C,review3d,review4,review5,Bamba:2012cp} and scenarios related to modified theories of gravity {\cite{Nojiri11,Clifton12}}.

The candidates of dark energy include cosmological constant and a
variety of scalar field models; the latter were invoked to alleviate
the problems associated with cosmological constant. Unfortunately, the scalar
field models are plagued with similar problems. The models based upon modified
theories of gravity are faced with challenges posed by the local physics.
 Large scale modification of gravity essentially involves extra degree(s) of
 freedom which might influence local physics where Einstein theory of gravity
 is in excellent agreement with observations. One then needs to invoke mass
 screening mechanisms to hide these degrees of freedom. To be fair, these
 scenarios do not perform better then the ones based upon dark energy.
 As for the latter,  one can reconstruct the cosmic history  referring to
  FRW background or by making use of the growth of perturbations on small
  scales. Given a priori a cosmic history specifying either the equation
  of state (EoS) or the scale factor $a$, one can always construct a
   scalar field potential which would mimic the desired result \cite{sf1,sf2,sf3,sf4,sf5,sf6,sf7,sf8}. Similar
    reconstruction can be carried out in scalar tensor theories.

On phenomenological grounds, a number of parametrization schemes
 have been investigated with the requirement of their theoretical consistency and  observational viability.
  In particular, parametrization of EoS/Hubble parameter/pressure have been extensively
  used in the literature \cite{staro,polar,linder,wparam}. The dynamics of realistic
  Universe is described by an EoS parameter which behaves differently at different epochs.
  For instance, in general relativistic description {of the dynamics of the
  spatially flat RW spacetime}, the fluids with constant EoS
  parameter $w>-1$ give rise to a power-law expansion ($a\propto t^{\frac{2}{3(1+w)}}$)
  of the Universe and to an exponential expansion $a\propto e^{k\,t}$, where $k>0$ is
   a constant, for $w=-1$. The solution of the Einstein's field equation in the presence
   of single fluid with a constant EoS parameter gives the relation between the EoS parameter
   of a fluid and the deceleration parameter (DP) of the Universe as,
    $q=-\frac{\ddot{a}a}{\dot{a}^2}=\frac{1+3w}{2}$. Obviously,
    a fluid with a constant EoS cannot give rise to realistic cosmic history.

The realistic  Universe should be dominated by stiff fluid {(which was suggested as the most
probable EoS for describing the very early Universe \cite{Zeldovich62,Barrow78})}, radiation,
pressureless matter and cosmological constant (or a fluid exhibiting a similar behavior with the
cosmological constant in the present Universe) respectively
as it evolves. In other words, the EoS parameter of the
effective fluid ($\rho_{\rm eff}=\sum \rho_{i}$) that
drives the expansion of Universe will not be yielding
a constant EoS, once we consider a mixture of fluids
with different EoS parameters at different epochs.

A variety of scalar field models including quintessence, K-essence,
 tachyons and phantoms investigated in the literature can give rise to
  a time dependent EoS parameter {\it a la} dynamical dark energy. Hence, the trajectory of the evolution of the DP would be dependent on the characteristics of the dark energy model. For instance, in most of the scalar field dark energy models, the effective EoS parameter of the dark energy evolves from $w=1$ to $w=-1$ so that the DP of the Universe evolves from $q=2$ to $q=-1$ in general relativity. However, different scalar field models would realize this by following different trajectories depending on the type of scalar field and the potential that describes the field.

In this paper we would like to investigate if a simple ansatz obtained by multiplying power-law and exponential law, which we call hybrid expansion law, could be successful in explaining the observed Universe. We point out the realization of such an expansion in Brans-Dicke theory in the presence of dust and also construct the scalar field models that can drive such an expansion law in the framework of general theory of relativity. We confront the model under consideration with the latest observational data and  discuss the results in the context of the late time cosmic acceleration. We further investigate the model with reference to Big Bang Nucleosynthesis (BBN), Baryonic Acoustic Oscillation (BAO) and Cosmic Microwave Background (CMB). Finally, we summarize the findings of the paper, and discuss the issues and future directions related to the ansatz considered in this study. 

\section{Hybrid expansion law in Robertson-Walker spacetime}
\label{sec:HEL}

In what follows, we shall consider the following ansatz referred to, hereafter, as hybrid expansion law (HEL):
\begin{equation}
\label{eqn:SF}
a(t)=a_{0}\left(\frac{t}{t_{0}}\right)^{\alpha}e^{\beta
\left(\frac{t}{t_{0}}-1\right)},
\end{equation}
where  $\alpha$ and $\beta$ are non-negative constants. Further, $a_{0}$ and $t_{0}$
respectively denote the scale factor and age of the Universe today.
The cosmological parameters; Hubble parameter, DP and jerk
parameter are respectively given by:
\begin{equation}
\label{eqn:HP}
H=\frac{\dot{a}}{a}=\frac{\alpha}{t}+\frac{\beta}{t_{0}},
\end{equation}
\begin{equation}
\label{eqn:DP}
q=-\frac{\ddot{a}}{aH^2}=\frac{\alpha t_{0}^2}{(\beta t+\alpha t_{0})^{2}}-1,
\end{equation}
\begin{equation}
\label{eqn:JP}
j=\frac{\dddot{a}}{aH^3}=1+\frac{ (2t_{0}-3{\beta} t-3{\alpha} t_{0})\alpha{t_{0}}^2}{(\beta t + \alpha t_{0})^3}.
\end{equation}

In particular cases, one obviously obtains
power-law and exponential expansion  from (\ref{eqn:SF}) choosing $\alpha=0$ and
$\beta=0$ respectively. It is evident that one may choose the constants such that the power-law term dominates over the exponential term in the early Universe, namely, at the time scales of the primordial nucleosynthesis ($t\sim 10^2\,{\rm second}$). Accordingly, for $t\sim 0$, the cosmological parameters approximate to the following:
\begin{equation}
\label{eqn:early}
a\sim a_{0}\left(\frac{t}{t_{0}}\right)^{\alpha},\quad H\sim \frac{\alpha}{t},\quad q\sim -1+\frac{1}{\alpha}\quad \textnormal{and}\quad j\sim 1-\frac{3}{\alpha}+\frac{2}{\alpha^2}.
\end{equation}
Similarly, the exponential term dominates at late times, such that in the limit $t\rightarrow \infty$ we have
\begin{equation}
\label{eqn:late}
a\rightarrow a_{0}e^{\beta\left(\frac{t}{t_{0}}-1\right)},\quad H\rightarrow \frac{\beta}{t_{0}},
\quad q\rightarrow -1 \quad\textnormal{and}\quad j\rightarrow 1.
\end{equation}

It may be observed that the parameter $\alpha$ determines the
initial kinematics of the Universe while the very late time
kinematics of the Universe is determined by the parameter $\beta$.
When $\alpha$ and $\beta$ both are non-zero,  Universe evolves with
variable DP given by \eqref{eqn:DP} and the transition from
deceleration to acceleration takes place at
$\frac{t}{t_{0}}=\frac{\sqrt{\alpha}-\alpha}{\beta}$,
which puts $\alpha$ in the range $0<\alpha<1$.

In the next section, we shall focus on HEL realization in the framework of Brans-Dicke theory while in the following section we shall study the corresponding effective fluid as well as single scalar field reconstruction in general theory of relativity. While doing these investigations, inspired by the inflationary scenarios, we shall consider only the spatially homogeneous and isotropic RW spacetime as the background geometry for describing the Universe.

\section{Hybrid expansion law from Brans-Dicke theory}
\label{sec:BD}

As it is demonstrated in Ref.\cite{Nojiri11}, it is always possible to carry out reconstruction program in the framework of scalar-tensor theory giving rise to a desired cosmic history. However, in what follows, we directly show that the particular case $\alpha=\frac{2}{3}$ of HEL is a particular solution of the Brans-Dicke field equations in the presence of dust fluid.

The action for the Jordan-Brans-Dicke (Brans-Dicke in Jordan frame) theory can be given as
\begin{eqnarray}
S= \int d^{4}x\sqrt{-g}\left[-\frac{\varphi^2}{8\omega}R+\frac{1}{2}\nabla_{\alpha}\varphi\nabla^{\alpha}\varphi\right]+S_{M},
\label{action}
\end{eqnarray}
where $S_{M}$ is the matter action, $\varphi$ is the Jordan field
{and $\omega$ is the Brans-Dicke coupling parameter/constant}.

The field equations obtained from this action for spatially flat RW
spacetime are as follows:
\begin{eqnarray}
3{\frac{\dot{a}^2}{a^2}}+6\frac{\dot{a}}{a}\frac{\dot{\varphi}}{\varphi}-2\omega
\frac{\dot{\varphi}^2}{\varphi^2}=\frac{4\omega}{\varphi^2}\rho,
\label{rho}
\\
\frac{\dot{a}^2}{a^2}+2\frac{\ddot{a}}{a}+2\frac{\ddot{\varphi}}{\varphi}+(2+2\omega)\frac{\dot{\varphi}^{2}}{\varphi^2}+4\frac{\dot{a}}{a}\frac{\dot{\varphi}}{\varphi}=-\frac{4\omega}{\varphi^2}p,
\label{pres}
\\
-\frac{3}{2\omega}\frac{\ddot{a}}{a}-\frac{3}{2\omega}\frac{\dot{a}^2}{a^2}+\frac{\ddot{\varphi}}{\varphi}+3\frac{\dot{a}}{a}\frac{\dot{\varphi}}{\varphi}=0,
\label{phi}
\end{eqnarray}
where $\rho$ and $p$ are the energy density and pressure of matter,
respectively. This system consists of three differential equations
(\ref{rho})-(\ref{phi}) that should be satisfied by four unknown
functions of $a$, $\varphi$, $\rho$ and $p$ and therefore is not
fully determined. At this point, considering the fact that cold dark
matter (CDM) and ordinary matter have zero pressure, we further assume
\begin{equation}
\frac{p}{\rho}=0,
\end{equation}
as an additional constraint to fully determine the system.

For the particular case $\omega=-\frac{4}{3}$, the solution of the
system is (with the adjustment $a(t)=0$ at $t=0$)
\begin{equation}
\label{eqn:BDsf}
a=\frac{c_{3}}{{c_{1}}^{\frac{2}{3}}}t^{\frac{2}{3}}\,e^{\frac{2}{3}c_{2}t},\quad \varphi=c_{1}t^{-\frac{1}{2}}\,e^{-c_{2}t}\quad\textnormal{and}\quad \rho=\frac{{c_{1}}^{2}\,c_{2}}{8}t^{-2}\,e^{-2c_{2}t},
\end{equation}
where $c_{1}$, $c_{2}$ and $c_{3}$ are integration constants. We
note that the scale factor we obtained here in \eqref{eqn:BDsf}
corresponds to the particular case $\alpha=\frac{2}{3}$ of our HEL
ansatz. It behaves as $a\sim t^{\frac{2}{3}}$ at $t\sim0$ and
evolves towards the exponential expansion monotonically with the passage of time
as in the $\Lambda$CDM cosmology, but follows a different
evolution trajectory. Thus the HEL, we used at the
beginning in ad hoc way, can be motivated by Brans-Dicke theory of
gravity. As a side remark, we notice that the values $|\omega | \sim
1$ of the Brans-Dicke parameter may be motivated by string theories,
namely, the low energy string effective action corresponds to
$\omega= -1$. Interestingly,  the value $\omega=-\frac{4}{3}$ that we used above for the particular
solution,  corresponds to the four-dimensional
spacetime with $0$-brane ($d=1$) in the $d$-branes string model
\cite{Duff95,Lidsey00}.

However, we should point out that Brans-Dicke theory with parameter
{$|\omega | \sim 1$} would  be in conflict with observations in the solar system. One then requires
to implement chameleon mechanism to satisfy local physics
constraints. One needs to enlarge the framework by invoking the
field potential such that the mass of the field gets heavy in high
density regime thereby escaping its detection locally. However, we shall not deal with these issues here.

\section{The effective fluid in general relativity}
%The special law for the expansion of Universe that we introduced in Section \ref{sec:HEL}, cannot be obtained by using a single fluid with a constant EoS parameter in Einstein gravity but might result due to the presence of certain mixture of different fluids or a single fluid with a time dependent EoS parameter.

\noindent The behavior of the scale factor under consideration in this paper was studied in the context of inflation in the early Universe by Parsons and Barrow in Ref. \cite{Parsons} (where a simple mathematical method for generating new inflationary solutions of Einstein's field equations from old ones was provided). They pointed out that the Einstein's field equations in the presence of self-interacting scalar field are invariant under constant rescaling of the scalar field, and then they generated the HEL behavior from power-law expansion. They also showed that such an expansion of the Universe can be represented as a Friedmann Universe in the presence of imperfect fluid that can be described by an EoS parameter of a perfect fluid with an added constant bulk viscous stress. In this paper, on the other hand, we study HEL expansion in the context of the history of the Universe after the inflation took place, and mainly investigate whether this law could be used for describing the evolution of the Universe starting from the radiation- or matter-dominated Universe to the currently accelerating Universe. Accordingly, we next discuss the effective fluid by interpreting it as a mixture of different sources that would lead to the HEL expansion in general relativity in Section \ref{sec:mixture}. We also study the single scalar field realization of the HEL expansion in Section \ref{sec:scalar}.

\subsection{Effective fluid as a mixture of different sources}
\label{sec:mixture}
In general relativity, one can always introduce an effective source that gives rise to a given expansion law. Accordingly, we obtain the energy density and EoS parameter of the effective source, which is assumed to describe the mixture of different types of sources such as matter, radiation and dark energy in general relativity. Hence, using the ansatz \eqref{eqn:SF} in the Friedmann equations in general relativity, we obtain the energy density and the EoS parameter of the effective fluid as follows:
\begin{equation}
\label{rhoweff}
\rho_{\rm eff}= 3\left(\frac{\alpha}{t}+\frac{\beta}{t_{0}}\right)^2\quad\textnormal{and}\quad w_{\rm eff}= \frac{2\alpha}{3t^2}\left(\frac{\alpha}{t}+\frac{\beta}{t_{0}}\right)^{-2}-1.
\end{equation}

The EoS parameter of the effective fluid starts with $w_{\rm eff} = -1+\frac{2}{3\alpha}$
at $t=0$, and evolves to $w_{\rm eff} \rightarrow {-1}$
as $t\rightarrow \infty$. One may observe that $\alpha$ is the parameter
that allows us to determine the effective fluid at early epochs.
Accordingly, choosing $\alpha=\frac{1}{3}$ we can set a beginning
with a stiff fluid domination, i.e., $w_{\rm eff} =1$ at $t\sim0$ for the
Universe, and choosing $\alpha=\frac{1}{2}$ we can set a beginning
with a radiation domination, i.e., $w_{\rm eff} =\frac{1}{3}$ at $t\sim0$.
Irrespective of the choice of the EoS parameter of the initial
effective fluid, the fluid evolves to the cosmological constant but
following different trajectories. Choosing $\alpha=\frac{1}{3}$, the
EoS parameter $w_{\rm eff} $ evolves from $1$ to $-1$, and the accelerated
expansion commences when
$\frac{t}{t_{0}}=\frac{1}{\sqrt{3}\,\beta}-\frac{1}{3\beta}$. Again choosing $\alpha=\frac{1}{2}$, the EoS parameter
$w_{\rm eff} $ evolves from $\frac{1}{3}$ to $-1$, and the accelerated
expansion starts at $\frac{t}{t_{0}}=\frac{1}{\sqrt{2}\,\beta}-\frac{1}{2\beta}$.

In this study, the value of $\alpha$ is not fixed to a certain value, but is left as a free parameter to be constrained using the latest data from $H(z)$ and SN Ia observations. Hence, the observational constraints on $\alpha$ shall determine the starting era of Universe within the framework of HEL model, while the $\Lambda$CDM model describes the Universe starting from the matter-dominated era. If it is found that $\alpha\sim\frac{1}{2}$, then the effective fluid may be interpreted as a mixture of radiation, matter and dark energy, and be written as $\rho_{\rm eff}=\rho_{\rm r}+\rho_{\rm m}+\rho_{\rm DE}$, where $\rho_{\rm r}\propto a^{-4}$ and $\rho_{\rm m}\propto a^{-3}$ stand for the radiation, matter (baryonic matter+CDM) constituents respectively, and $\rho_{\rm DE}$ stands for the unknown dark energy source that gives rise to HEL expansion together with radiation and matter. In this case, HEL model may be considered as a candidate for describing the Universe starting from the radiation dominated era, and hence one can further investigate the HEL model by discussing the primordial nucleosynthesis times, and checking whether the matter-dominated era would be recovered properly or not. On the other hand, if it is found that $\alpha\sim\frac{3}{2}$,  then the effective fluid may be interpreted as a mixture of matter and dark energy, and be written as $\rho_{\rm eff}=\rho_{\rm m}+\rho_{\rm DE}$, like in the $\Lambda$CDM model for which dark energy is given by cosmological constant $\Lambda$. Hence, in this case, HEL model may be considered as a candidate for describing the Universe starting from the matter-dominated era as in the $\Lambda$CDM model.

We note here that the dark energy fluid can be obtained by subtracting the known constituents (such as matter and radiation) from the effective fluid.  We shall stick to this approach in our discussions that follow the observational analysis. On the other hand, it might be interesting and useful to see the single scalar field correspondence of the HEL expansion before doing the observational analysis.

\subsection{Effective fluid as a single scalar field}
\label{sec:scalar}
We can always construct a scalar field Lagrangian which can mimic a given expansion law. Accordingly, in this section, assuming that the effective energy density and EoS parameter given in (\ref{rhoweff}) correspond to a single scalar field, we obtain the potentials for the quintessence, tacyhon and phantom fields, which are the most commonly considered scalar field candidates for dark energy. One can use these potentials for describing dark energy sources.

\subsubsection{Quintessence field correspondence}

Most of the dark energy studies are carried out within the quintessence paradigm of a slowly rolling canonical scalar field with a potential. Therefore, we first consider the quintessence realization of the HEL. The energy density and pressure of the quintessence minimally coupled to gravity can be given by
\begin{equation}
\rho =\frac{1}{2}\dot{\phi}^2+V(\phi)
\quad\textnormal{and}\quad
p =\frac{1}{2}\dot{\phi}^2-V(\phi),
\end{equation}
where $\phi$ is the canonical scalar field with a potential $V(\phi)$. Using these with the HEL ansatz \eqref{eqn:SF}, we obtain
\begin{equation}
\phi(t)=\sqrt{2\alpha}\ln(t)+\phi_{1}
\quad\textnormal{and}\quad
V{(t)}=3\left(\frac{\alpha}{t}+\frac{\beta}{t_{0}}\right)^2-\frac{\alpha}{t^2},
\end{equation}
where $\phi_{1}$ is the integration constant. The potential as a function of the scalar field $\phi$ is then given by the following expression:
\begin{equation}
V(\phi)=3\beta^{2}e^{-\sqrt{\frac{2}{\alpha }}(\phi_{0}-\phi_{1})}+\alpha(3\alpha-1)e^{-\sqrt{\frac{2}{\alpha }}(\phi-\phi_{1})}+6\alpha\beta e^{-\frac{1}{2}\sqrt{\frac{2}{\alpha }}(\phi+\phi_{0}-2\phi_{1})},
\end{equation}
where $\phi_{0}=\phi_{1}+\sqrt{2\alpha }\ln(t_{0})$.

One may observe that this potential can be seen as the summation of three different potentials, i.e., a constant potential and two exponential potentials. Choosing $\alpha=0$ the potential reduces to a constant, hence to a cosmological constant, that would give rise to exponential expansion. Choosing $\beta=0$, the potential reduces to a single exponential potential that would give rise to power-law expansion (see for instance \cite{Gumjudpai12}) and may describe a matter field with a constant EoS parameter. For $\alpha\neq 0 \neq \beta$, on the other hand, the potential contains a constant on the left, an exponential potential in the middle and an additional potential term depending on both $\alpha$ and $\beta$, which may be interpreted as an interaction term between the first two potentials.

One may observe that the condition for non-negativity of the potential is $\alpha\geq \frac{1}{3}$. Under this condition, we have
$
\frac{1}{2}\dot{\phi}^2\rightarrow \infty \quad\textnormal{and}\quad V\rightarrow \infty\quad \textnormal{as } t\rightarrow 0
$,
and
$
\frac{1}{2}\dot{\phi}^2\rightarrow 0 \quad\textnormal{and}\quad V\rightarrow 3\beta^{2} e^{-\sqrt{\frac{2}{\alpha }}(\phi_{0}-\phi_{1})}\quad \textnormal{as } t\rightarrow \infty.
$
We see that the scalar field approaches  positive cosmological constant at late times of the Universe.

\subsubsection{Tachyon field correspondence}
Quintessence paradigm relies on the potential energy of scalar fields to drive the late time acceleration of the Universe. On the other hand, it is also possible to relate the late time acceleration of the Universe with the kinetic term of the scalar field by relaxing its canonical kinetic term. This idea is known as k-essence \cite{Picon00}. In this section we are interested in a special case of k-essence that is known as Tachyon. Tachyon fields can be taken as a particular case of k-essence models with Dirac-Born-Infeld (DBI) action and can also be motivated by string theory \cite{Gibbons02,Chimento04}. It has been of interest to the dark energy studies due to its EoS parameter $w=\dot\phi^2-1$ that evolves smoothly from 0 to -1 \cite{Bagla03,Copeland05,Calcagni06}.

In case of Tachyon field, the energy density and pressure read as
\begin{equation}
\rho =\frac{V(\phi)}{\sqrt{1-{\dot\phi}^2}}
\quad\textnormal{and}\quad
p =-V(\phi)\sqrt{1-{\dot\phi}^2},
\end{equation}
where $\phi$ is the Tachyon field with potential $V(\phi)$. Using these with the HEL ansatz \eqref{eqn:SF}, we find
\begin{equation}
\phi(t)=\sqrt{\frac{2\alpha t_{0}^2}{3\beta^{2}}}\ln(\beta t+\alpha t_{0})+\phi_{2}
\quad\textnormal{and}\quad
V{(t)}=3\left(\frac{\alpha}{t}+\frac{\beta}{t_{0}}\right)^2\sqrt{1-\frac{2\alpha t_{0}^2}{3(\beta t+\alpha t_{0})^2}},
\end{equation}
where $\phi_{2}$ is an integration constant. The tachyon potential that drives the HEL Universe is given by
\begin{equation}
\label{eqn:tpot}
V(\phi)=\frac{3 \beta ^2}{t_{0}^2}e^{\sqrt{\frac{6\beta^2}{\alpha t_{0}^2}}(\phi-\phi_{2})}\sqrt{1-\frac{2}{3}\alpha t_{0}^2 e^{\sqrt{\frac{6\beta^2}{\alpha t_{0}^2}}(\phi-\phi_{2})}}\left(\alpha t_{0}-e^{\frac{1}{2}\sqrt{\frac{6\beta^2}{\alpha t_{0}^2}}(\phi-\phi_{2})}\right)^{-2}.
\end{equation}

We see that the tachyon potential is real subject to the condition $\alpha\geq \frac{2}{3}$. Also the EoS parameter $w=\dot\phi^2-1$ in this case varies from $0$ to $-1$, as it should be for the tachyon field, during the evolution of Universe.

\subsubsection{Phantom field correspondence}
Quintessence and tacyhon fields considered in the previous two subsections can yield EoS paremeters $w\geq -1$. However, the observations at present allow slight phantom values for the EoS parameter, i.e., $w<-1$ \cite{Vazquez12,Nobosyadlyj12,Parkinson12,Ade13}. In case, this is confirmed by future observations, the latter might have far reaching consequences for the fate of Universe. It is thus interesting to examine the phantom dynamics. Sources behaving as a phantom field can arise in braneworlds, Brans-Dicke scalar-tensor gravity and may be motivated from $S$-brane constructions in string theory \cite{Shtanov03,Elizade04,Chen02,Townsent03,Ohta03}. On the other hand, the phantom energy, in general, can be simply described by a scalar field with a potential $V(\phi)$ like the quintessence dark energy but with a negative kinetic term \cite{Caldwell02}. Accordingly, the energy density and pressure of the phantom field are respectively given by
\begin{equation}
\rho =-\frac{1}{2}\dot{\phi}^2+V(\phi)\quad{\rm and}\quad p =-\frac{1}{2}\dot{\phi}^2-V(\phi),
\end{equation}
where $\phi$ is the phantom field with potential $V(\phi)$.

In case of the phantom scenario, the HEL ansatz (1) must be slightly modified in order to acquire self consistency. In particular, we rescale time as $t\rightarrow t_{s}-t$, where $t_{s}$ is a sufficiently positive reference time. Thus, the HEL ansatz \eqref{eqn:SF}  becomes
\begin{equation}
a(t)=a_{0}\left(\frac{t_{s}-t}{t_{s}-t_{0}}\right)^{\alpha}e^{\beta \left(\frac{t_{s}-t}{t_{s}-t_{0}}-1\right)},
\end{equation}
and the Hubble parameter, its time derivative and DP read as
\begin{equation}
H=-\frac{\alpha}{t_{s}-t}-\frac{\beta}{t_{s}-t_{0}},
\quad
\dot{H}=-\frac{\alpha}{(t_{s}-t)^2}
\quad\textnormal{and}\quad
q=\frac{\alpha (t_{s}-t_{0})^2}{[\beta (t_{s}-t)+\alpha (t_{s}-t_{0})]^{2}}-1.
\end{equation}
We observe that $\alpha<0$ leads to $q<0$ (acceleration) and $\dot{H}>0$ (super acceleration). Also the scale factor and Hubble parameter diverge as $t\rightarrow t_{s}$ and thus exposing the Universe to Big Rip. Further, we find
\begin{equation}
\phi(t)=\sqrt{-2\alpha }\ln(t_{s}-t)+\phi_{3}
\quad\textnormal{and}\quad
V{(t)}=3\left(\frac{\alpha}{t_{s}-t}+\frac{\beta}{t_{s}-t_{0}}\right)^2-\frac{\alpha}{(t_{s}-t)^2},
\end{equation}
where $\phi_{3}$ is an integration constant. The phantom potential in terms of the phantom field reads as
\begin{equation}
V(\phi)=3\beta^{2}e^{-\sqrt{-\frac{2}{\alpha }}(\phi_{0}-\phi_{3})}+\alpha(3\alpha-1)e^{-\sqrt{-\frac{2}{\alpha }}(\phi-\phi_{3})}+6\alpha\beta e^{\frac{1}{2}\sqrt{-\frac{2}{\alpha }}(\phi+\phi_{0}-2\phi_{3})},
\end{equation}
where $\phi_{0}=\phi_{3}+\sqrt{-2\alpha }\ln(t_{s}-t_{0})$. In the phantom HEL cosmology, we find that at late times $w=-1+\frac{2}{3\alpha}$, which lies below the phantom divide line ($w=-1$) for $\alpha<0$ as expected.

\bigskip
The aforesaid discussion shows that HEL finds simple expressions in models of scalar fields and provides a simple way of transition from acceleration to deceleration. In case of phantom field alone, we are in the phase of super acceleration and cannot cross the phantom divide line ($w=-1$) in a single field model. We then need to tune phantom field such that it remains sub-dominant during matter phase and shows up only at late times. In case of quintessence and tachyon, the field can formally mimic dark matter like behavior which is impossible in case of a phantom field.

\section{Observational constraints on HEL cosmology from $H(z)$+SN Ia data}
\label{sec:obs}

For observational purposes, we use the following relation between scale factor and redshift:
\begin{equation}
a(t)=\frac{a_0}{1+z}.
\end{equation}

Invoking HEL \eqref{eqn:SF} into the above equation and solving for $t$, we have
\begin{equation}
t=\frac{\alpha t_0}{\beta}f(z),
\end{equation}
where
\begin{equation}
f(z)=\,{\rm LambertW} \left( \frac{\beta}{\alpha}\,{{\rm e}^{{\frac {\beta-\ln  \left(1+z \right) }{\alpha}}}} \right).\nonumber
\end{equation}

The Hubble parameter in terms of redshift for the HEL cosmology reads as
\begin{equation}
H(z)=\frac{H_0\beta}{\alpha+\beta}\left[\frac{1}{f(z)}+1\right],
\end{equation}
where $H_0=\frac{\alpha+\beta}{t_0}$. We see that the parameter space of HEL cosmology consists of three parameters namely $\alpha$, $\beta$ and $H_0$ to be constrained by the observations. 

The authors of Ref.\cite{ref:SVJ2005} obtained nine $H(z)$ data points from the relative dating of 32 passively evolving galaxies. Using the BAO
peak position as a standard ruler in the radial direction, $H(z)$ was estimated for $z=0.24$ and $z=0.45$ in Ref.\cite{Gaztanaga09}. Two determinations for $H(z)$ were given in Ref.\cite{stern10} using red-envelope galaxies while a reliable sample of eight $H(z)$ points was derived in Ref.\cite{moresco12} using the differential spectroscopic evolution of early-type galaxies
as a function of redshift. The authors of Ref.\cite{ref:Zhang2012}  presented four $H(z)$ points adopting the differential age method and utilizing selected 17832 luminous red galaxies from Sloan Digital Sky Survey (SDSS) Data Release Seven. We compile all the 25 $H(z)$ data points spanning in the redshift
range $0.07 < z < 1.750$ in Table  \ref{Table:1a}. It may, however, be noted that though the  $H(z)$ data points derived from different methods/sources are frequently used in the literature for constraining cosmological parameters but these are prone to systematics. Following the methodology given in Ref.\cite{lvdp13}, we utilize these 25 observational $H(z)$ data points in addition to the SN Ia Union2.1 sample \cite{ref:Suzuki2012} that contains 580 SN Ia data points spanning in the redshift range $0.015 < z < 1.414$ for constraining the parameters of HEL cosmology. We use the Markov Chain Monte Carlo (MCMC) method, whose code is based on the publicly available package {\bf cosmoMC} \cite{ref:MCMC}, for the data analyses. We have also constrained the standard $\Lambda$CDM model parameters with the same observational data sets of $H(z)$ and SN Ia for the sake of comparison with the HEL models {(See Appendix \ref{sec:LCDM} for the dynamics of the $\Lambda$CDM cosmology)}.

\begin{center}
\begin{table}\small\centering\footnotesize\caption{$H(z)({\rm km}\, {\rm s}^{-1}\,{\rm Mpc}^{-1})$ measurements with $1\sigma$ errors.}\label{Table:1a}
\begin{tabular}{lllc}
\hline\hline
$z\quad\quad\quad\quad$ & $H(z)\quad\quad$  & $\sigma_{H(z)}$ & Reference \\\hline\hline
0.090  & 69    & 12    & \cite{ref:SVJ2005} \\
    0.170 & 83    & 8     & \cite{ref:SVJ2005} \\
    0.270 & 77    & 14    & \cite{ref:SVJ2005} \\
    0.400  & 95    & 17    & \cite{ref:SVJ2005} \\
    0.900 & 117   & 23    & \cite{ref:SVJ2005} \\
    1.300 & 168   & 17    & \cite{ref:SVJ2005} \\
    1.430 & 177   & 18    & \cite{ref:SVJ2005} \\
    1.530 & 140   & 14    & \cite{ref:SVJ2005} \\
    1.750 & 202   & 40    & \cite{ref:SVJ2005} \\
    0.24  & 79.69  &3.32  & \cite{Gaztanaga09}\\
    0.43  &86.45  &3.27  & \cite{Gaztanaga09}\\
    0.480 & 97    & 62    & \cite{stern10} \\
    0.880 & 90    & 40    & \cite{stern10} \\
    0.179 & 75    & 4     & \cite{moresco12} \\
    0.199 & 75    & 5     & \cite{moresco12} \\
    0.352 & 83    & 14    & \cite{moresco12} \\
    0.593 & 104   & 13    & \cite{moresco12} \\
    0.680 & 92    & 8     & \cite{moresco12}\\
    0.781 & 105   & 12    & \cite{moresco12} \\
    0.875 & 125   & 17    & \cite{moresco12} \\
    1.037 & 154   & 20    & \cite{moresco12} \\
0.07  & 69.0  & 19.6 & \cite{ref:Zhang2012} \\
0.12  & 68.6  & 26.2 & \cite{ref:Zhang2012}   \\
0.20  & 72.9  & 29.6 & \cite{ref:Zhang2012} \\
0.28  & 88.8  & 36.6 & \cite{ref:Zhang2012}  \\
\hline\hline\\
\end{tabular}
\end{table}
\end{center}

The 1D marginalized distribution on individual parameters and 2D contours with 68.3 \%, 95.4 \% and 99.73 \% confidence limits are shown in Fig.\ref{fig:LVDPzBR} for the HEL model. The mean values of the HEL model parameters $\alpha$, $\beta$ and $H_{0}$ constrained with $H(z)$+SN Ia data are given in Table \ref{table:HELobs}. The $1\sigma$, $2\sigma$ and $3\sigma$ errors, $\chi^2_{min}$ and $\chi^2_{min}$/dof are also given in Table \ref{table:HELobs}.
\begin{figure}[htb!]\centering
\psfrag{q0}[b][b]{$t$}
%\psfrag{z}[b][b]{$z$}
\includegraphics[width=12cm]{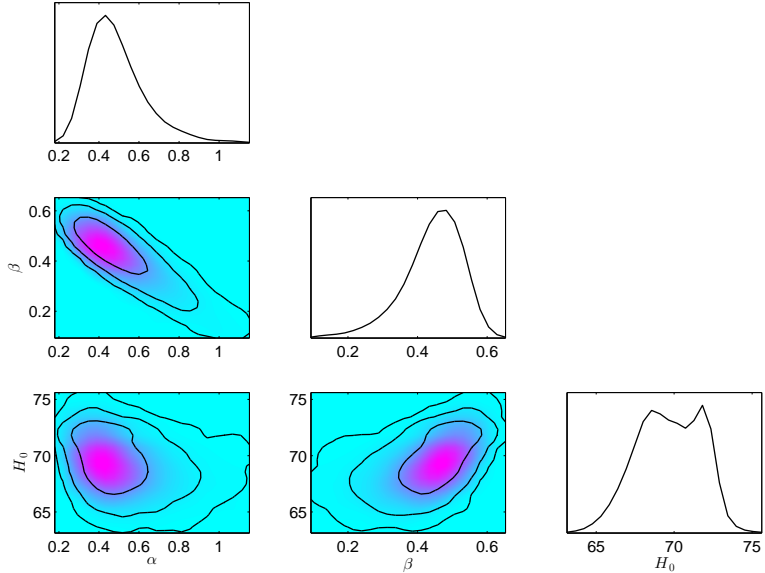}
\caption{\footnotesize{The 1D marginalized distribution on individual parameters of HEL model and 2D contours with 68.3 \%, 95.4 \% and 99.7 \% confidence levels are obtained by using $H(z)$+SN Ia data points. The shaded regions show the mean likelihood of the samples.}}
\label{fig:LVDPzBR}
\end{figure}
\begin{table}[htb!]\centering\small
\caption{Mean values with errors of the HEL model parameters constrained with $H(z)$+SN Ia data.}
\begin{tabular}{ll}
\hline\hline Parameters & Mean values with errors  \\ \hline\hline\\
$\alpha$ & $    0.488_{-    0.128-    0.196-    0.260}^{+    0.124+    0.353+    0.602}$ \\[8pt]
$\beta$ & $    0.444_{-    0.077-    0.204-    0.323}^{+    0.079+    0.127+    0.173}$ \\[8pt]
$H_0$ (${\rm km}\,s^{-1}\,{\rm Mpc}^{-1}$)& $   69.682_{-    2.090-    3.979-    5.820}^{+    2.316+    3.209+    5.050}$ \\[8pt]
$\chi^2_{min}$ & $557.161$ \\[8pt]
$\chi^2_{min}$/dof & $0.9209$ \\[8pt]
\hline\hline
\end{tabular}
\label{table:HELobs}
\end{table}

In Table \ref{table:HELLCDM1}, we give the values of the cosmological parameters with the $1\sigma$ errors that we obtained using the HEL and $\Lambda$CDM models, viz., age of the present Universe $t_{0}$;  Hubble constant $H_{0}$, current values of the DP $q_{0}$ and jerk parameter $j_{0}$,  time passed since the accelerating expansion started $t_{0}-t_{\rm tr}$, redshift of the onset of the accelerating expansion $z_{\rm tr}$, energy density $\rho_0$ and the EoS parameter $w_{0}$ of the effective fluid at the present epoch of evolution of the Universe. We also give $\chi^2_{min}$ and $\chi^2_{min}$/dof in Table \ref{table:HELLCDM1} to compare the success of the models on fitting the data. We notice that both the models fit  observational data with a great success but the $\Lambda$CDM model fits the data slightly better than the HEL model does.
\begin{table}[htb!]\centering\small
\caption{Mean values with 1$\sigma$ errors of some important cosmological parameters  related to HEL and $\Lambda$CDM models. $\chi^2_{min}$/dof, AIC, KIC and BIC values are also displayed.}
\begin{tabular}{llll}
\hline\hline Parameters & HEL & $\Lambda$CDM \\ \hline\hline\\
$t_0 $ (Gyr)& $13.078\pm 1.096$ & $13.389 \pm  0.289$ \\[8pt]
$H_0$ (${\rm km}\,s^{-1}\,{\rm Mpc}^{-1}$) & $69.682_{-2.090}^{+2.316}$ & $70.697_{-2.020}^{+1.667}$ \\[8pt]
$q_0$ & $-0.438\pm0.094$ & $-0.556\pm 0.046$  \\[8pt]
$j_{0}$ & $0.520\pm 0.156$ & $1$ \\[8pt]
 $t_{0}-t_{\rm tr}$ (Gyr) & $6.874\pm1.558$ & $6.156\pm 0.366$  \\[8pt]
 $z_{\rm tr}$ & $0.817_{-0.141}^{+0.394}$  & $0.682\pm 0.082$ \\[8pt]
  $\rho_{0}\;(10^{-27}$ kg m$^{-3}$) & $9.122\pm 0.528$ & $9.389\pm 0.481$ \\[8pt]
$w_{0}$ & $-0.625\pm 0.063$ & $-0.704\pm 0.030$\\[8pt]\hline
$\chi^2_{min}$ & $557.161$ & $556.499$  \\[8pt]
$\chi^2_{min}$/dof & $0.9209$ & $0.9198$ \\[8pt]
AIC & $563.16$   & $560.50$     \\[8pt]
KIC &$566.16 $     & $562.50 $   \\[8pt]
BIC & $576.37$   &$569.31$\\[8pt]
\hline\hline
\end{tabular}
\label{table:HELLCDM1}
\end{table}

Next, for comparing the $\Lambda$CDM and HEL models we make use of statistical tools such as Akaike Information Criterion (AIC), Kullback Information Criterion (KIC) and Bayes Information Criterion (BIC) , which are commonly used in modern cosmology for model selection among competing models.  For instance, in a recent paper \cite{Melia2013}, Melia and Maier used these information criteria to compare $\Lambda$CDM model and $R_h=ct$ Universe. The three information criteria are defined as follows (see Ref.\cite{Melia2013} and references therein for details): 
\begin{equation}
\nonumber
\text{AIC}=\chi^2+2k,\; \text{KIC}=\chi^2+3k,\;\text{BIC}=\chi^2+k\ln n\;,
\end{equation}
where $k$ is number of model parameters and $n$ is number of data used in fitting.
A model with lower value of AIC, KIC or BIC is considered to be closest to the real model. So these information criteria provide relative evidence of better model among the models under consideration. Further, the difference of the AIC values of two models is denoted by $\Delta \text{AIC}$. A rule of thumb used in the literature is that if $\Delta \text{AIC}\lesssim 2$, the evidence is weak; if $\Delta \text{AIC}\approx 3 \;\text{or}\; \approx 4$, it is mildly strong and in case $\Delta \text{AIC}\gtrsim 5$, it is quite strong. Similar rule of thumb is followed for testing the strength of evidence while dealing with KIC. In case of BIC, the evidence is judged positive for the values  of $\Delta \text{BIC}$ between 2 and 6. A value of greater than 6 indicates strong evidence.
In the case in hand, we have two models namely $\Lambda$CDM and HEL. The corresponding values of AIC, KIC and BIC are given in Table \ref{table:HELLCDM1}. We immediately find that $\Delta$AIC $= 2.66$, $\Delta$KIC $= 3.66$ and $\Delta$BIC $= 7.06$. These figures from $H(z)$+SN Ia data only suggest that $\Lambda$CDM model is favored over the HEL model. One may see that AIC does not offer a strong evidence against the HEL model. However, the HEL model pays penalty in KIC and BIC cases because it carries one additional parameter in comparison to the $\Lambda$CDM model.

We note
all the cosmological parameters related with the present day
Universe as well as with the onset of the acceleration given in
Table \ref{table:HELLCDM1} for the HEL and $\Lambda$CDM models are
consistent within the $1\sigma$ confidence level. Only exception is that
the present values of the jerk parameter do not coincide in the two
models within the $1\sigma$ confidence level. However, we should
recall that jerk parameter is determined to be a constant
$j_{\Lambda{\rm CDM}}=1$ in $\Lambda$CDM, and hence doesn't involve
error, while it is function of time with two free parameters
$\alpha$ and $\beta$ in the HEL model. We additionally note that jerk
parameter involves the third time derivative of the scale factor, and
consequently it is constrained observationally rather weakly
\cite{sahni03,Visser04,Cattoen08,Wang09,Vitagliano10,Capozziello11,Xia12}.
Hence, we are not able to decide which model describes the expansion
of the Universe well considering the jerk parameter. It is, on the other
hand, a very useful parameter to see how the HEL model deviates from
the $\Lambda$CDM model. In accordance with this at the end of this
section, we shall also compare the evolution trajectories of the HEL
and $\Lambda$CDM models in the plane of DP and jerk parameter.

\begin{table*}[htb!]\small
\caption{Mean values and asymptotic limits with 1$\sigma$ errors of various parameters pertaining to the HEL and $\Lambda$CDM models. }
\centering
\begin{tabular}{|c|c|c|c|c|c|c|c|c|c|}
\hline
Model $\rightarrow$ & \multicolumn{3}{c|}{HEL} &  \multicolumn{3}{c|}{$\Lambda$CDM}\\\hline
Parameter & $z\rightarrow \infty$ & $z=0$ & $z\rightarrow -1$ & $z\rightarrow \infty$ & $z=0$&$z\rightarrow -1$\\\hline
$H\;$(km s$^{-1}$ Mpc$^{-1}$) & $\infty$ & $69.682_{-2.090}^{+2.316}$ & $33.164\pm 8.604$ & $\infty$ & $70.697_{-2.020}^{+1.667}$ & $59.336\pm 2.592$ \\[6pt]
$q$ & $1.049\pm 0.590$& $-0.438\pm0.094$ & $-1$& $0.5$ & $-0.556\pm 0.046$ & $-1$\\[6pt]
$j$ & $3.240\pm 3.060$ & $0.520\pm 0.156$ & $1$ & $1$& $1$ &$1$\\[6pt]
$\rho\;(10^{-27}$ kg m$^{-3}$) & $\infty$ & $9.122\pm 0.528$ & $2.066\pm 1.072$ & $\infty$ & $9.389\pm 0.481$ & $6.614\pm 0.577$\\[6pt]
$w$ & $0.364\pm 0.393$ & $-0.625\pm 0.063$ & $-1$ & $0$ & $-0.704\pm 0.030$ & $-1$ \\[6pt]
\hline
\end{tabular}
\label{table:HELLCDM2}
\end{table*}

In Table \ref{table:HELLCDM2}, we give the values of some important cosmological parameters with 1$\sigma$ errors for
the HEL and $\Lambda$CDM models at three different epochs:
early epoch ($z\rightarrow\infty$), present epoch ($z=0$) and future epoch ($z\rightarrow -1$). One may see that both the models exhibit similar behaviors at the present and future epochs. Regarding past of the Universe in the $\Lambda$CDM model, we emphasize that the $z\rightarrow\infty$ limit of the $\Lambda$CDM model is already determined as the dust dominated Universe in general relativity, and in fact this model can be used for describing the actual Universe for redshift values less than $z\sim3400$. On the other hand, it is interesting that the predicted early Universe ($q_{z\rightarrow\infty}\sim 1$ and $w_{z\rightarrow\infty}\sim\frac{1}{3}$)  in the HEL model is a very good approximation to the radiation dominated Universe in general relativity (where, $q=1$ and $w=\frac{1}{3}$). This motivates us to further investigate the early Universe behavior  in the HEL model, that we shall do in the next section.

%%%%%%%%%%%%%%%%%%%%

In the aforesaid, we  observed and discussed the particular values  of the
cosmological parameters in HEL and
 $\Lambda$CDM models for the present
 Universe and for two extremes
  $z\rightarrow\infty$ and $z\rightarrow -1$.
  Next, we would like to conclude this section by comparing the
  continuous evolution of these models. A very useful
  way of comparing and distinguishing different
  cosmological models, that have similar kinematics,
  is to plot the evolution trajectories of
  the $\{q,j\}$ and $\{j,s\}$ pairs. Here,
  $q$ and $j$ have the usual meaning and
  $s$ is a parameter defined as
\begin{equation}
\label{eqn:sfder}
s=\frac{j-1}{3(q-\frac{3}{2})}.
\end{equation}
In the above definition $s$, there is $\frac{3}{2}$ in the place of $\frac{1}{2}$ in the original definition $s=\frac{j-1}{3(q-\frac{1}{2})}$ by Sahni et al. \cite{sahni03}. This is to avoid the divergence of the parameter $s$ when the HEL model passes through $q=1$ or $q=\frac{1}{2}$ as done in Ref. \cite{lvdp13}. The parameter $s$ was originally introduced to characterize the properties of dark energy, and hence the evolution of the Universe was considered starting from dust dominated era in general relativity, which gives $q=\frac{1}{2}$ and $j=1$. However, in accordance with the HEL model, here we are also interested in the possibility of describing the Universe starting from primordial nucleosynthesis times, where the expansion of the Universe can be best described by $q\sim 1$. Accordingly, using (\ref{eqn:DP}) and (\ref{eqn:JP}) in (\ref{eqn:sfder}) we get
\begin{eqnarray}
s=\frac{2 \alpha t_0^2 [3 \beta t+(3 \alpha -2) t_0]}{3 (\beta t+\alpha t_0) \left[5 (\beta t+\alpha t_0)^2-2\alpha t_{0}^2\right]}
\end{eqnarray}
for the HEL model.

We plot evolution trajectories of the HEL and $\Lambda$CDM models in the $j-q$ plane in Fig. \ref{fig:rq} and in the $j-s$ plane in Fig. \ref{fig:rs} in the range $-1\leq q<1$ by considering the mean values of the model parameters given in Table \ref{table:HELLCDM1} from observations.

For comparison, we include also some alternatives to the $\Lambda$CDM model such as the Galileon, Chaplygin Gas and DGP models (for these models see \cite{Sami12} and references therein) in the figures. The arrows on the curves show the direction of evolution and the dots on the curves represent the present values of the corresponding $\{q,j\}$ and $\{j,s\}$ pairs while the black dots show the matter dominated phases of the models.

\begin{figure}[htb!]
\centering
\subfigure[]
{\psfrag{r1}[b][b]{$j$}
\psfrag{q3}[b][b]{$q$}
\includegraphics[width=7cm]{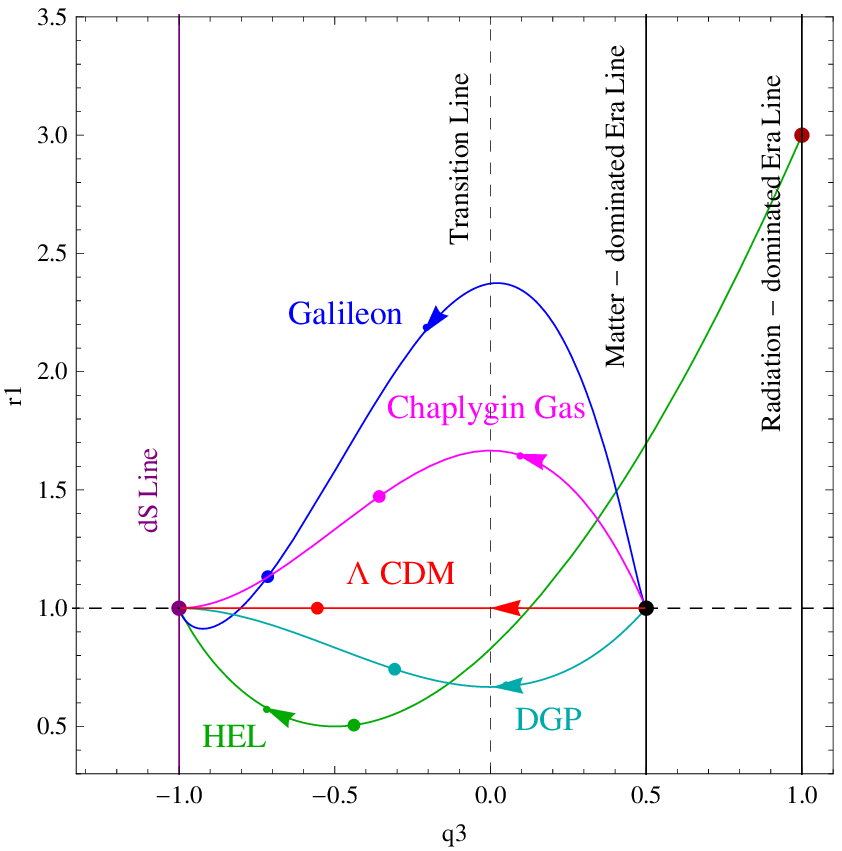} \label{fig:rq}}
\hspace{0.1cm}
\subfigure[]{
\psfrag{r1}[b][b]{$j$}
                \psfrag{s1}[b][b]{$s$}
\includegraphics[width=7 cm]{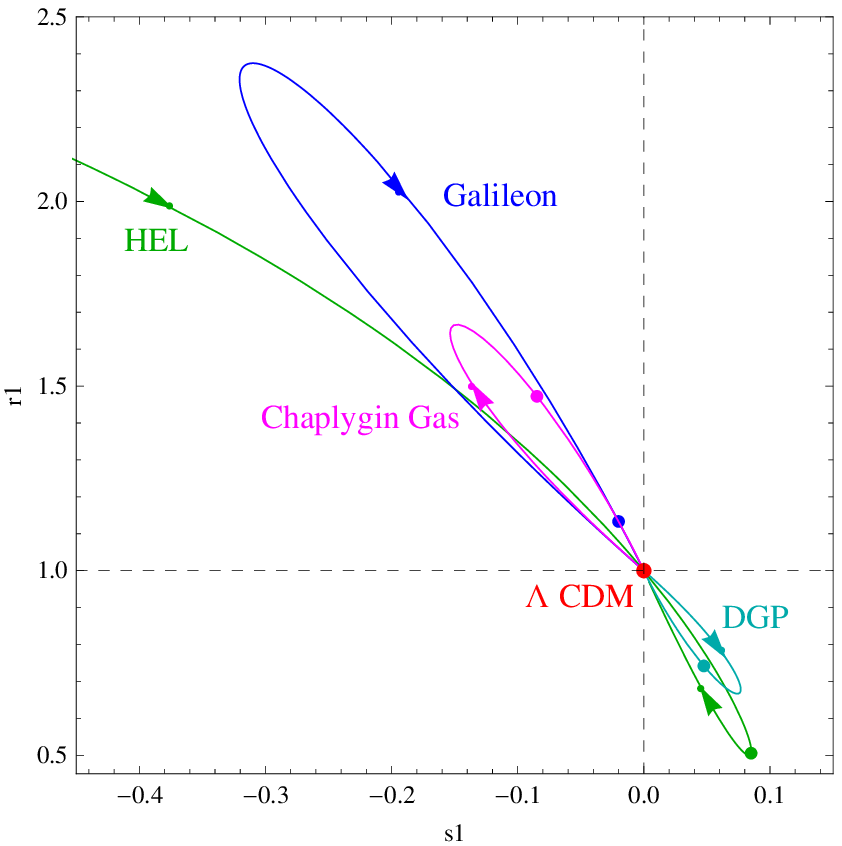} \label{fig:rs}}
\caption{\footnotesize{ \textbf{(a)} Variation of $q$ versus $j$. Vertical Purple line stands for the de Sitter (dS) state $q=-1$. \textbf{(b)} Variation of $s$ versus $j$.  Horizontal and vertical dashed lines intersect at the $\Lambda$CDM point $(0,1)$. In both panels, the Green curve corresponds to the HEL model.  Red, Cyan, Magenta and Blue curves correspond to $\Lambda$CDM, DGP, Chaplygin gas and Galileon models respectively. The arrows on the curves show the direction of evolution. The dots on the curves represent the present values of the corresponding $(s,j)$ or $(q,j)$ pair while the black dots show the matter dominated phases of the models. The dark red dot on the radiation-dominated line  corresponds to the HEL model. Thus, it starts from the radiation-dominated phase and evolves to de Sitter phase while all other models under consideration evolve from the SCDM phase to the de Sitter phase.}}
\end{figure}

We observe that all the models have different evolution trajectories
but the values of $q$, $j$ and $s$ do not deviate a lot in different
models for {$q\lesssim 0.5$} and are destined to the same future (de Sitter Universe). We note that only in the HEL model, the
actual Universe can be described down to the BBN times of the
Universe. In Figure \ref{fig:rs}, we observe that all trajectories except the one related to HEL
model are closed curves, namely they start and end at
point $\{j,s\}=\{1,0\}$. In the HEL model, on the other hand, the
evolution trajectory does not start at $\{j,s\}=\{1,0\}$.

%%%%%%%%%%%%%%%%%%%%%%%%%%%%
\section{Further investigations of the HEL model}

In the following, using the values of the model parameters obtained from the $25+580$ data points from the latest $H(z)$ and SN Ia compilations spanning in the redshift range $0.015<z<1.750$ in the previous section, we discuss whether the HEL model makes successful predictions for high redshift values ($z\sim 10^{9}-10^{8}$) considering BBN in Section \ref{sec:early}, for low redshift values ($z\sim 0$) considering BAO in Section \ref{sec:today} and then for intermediate redshift values considering, particularly, CMB in Section \ref{sec:intermediate}. 

\subsection{BBN test }
\label{sec:early}
It is showed in Section \ref{sec:obs} that HEL law predicts the value of the DP at $z\rightarrow\infty$, i.e., in the early Universe, as $q_{z\rightarrow\infty}=1.049 \pm 0.590$ ($1\sigma$), which can be maintained by the presence of an effective fluid that yields an EoS parameter $w_{z\rightarrow\infty}=0.364\pm0.393$ in general relativity. It is interesting that this predicted early Universe in the HEL model using the cosmological data related with the present day Universe is in good agreement with our conventional
expectations on the early Universe, viz., it should have been dominated by radiation ($w\sim\frac{1}{3}$) and expanding with a DP $q\sim1$. Hence, we first discuss the early Universe prediction in the HEL model which can be done through the BBN processes that occur at redshift range $z\sim10^{9}-10^{8}$ (when the temperature ranges from $T\sim 1\,{\rm MeV}$ to $T\sim 0.1\, {\rm MeV}$ and the age of the Universe varies from $t\sim 1\,{\rm s}$ to $t\sim 3\,{\rm min}$).

$^4$He mass ratio $Y_{\rm p}\equiv\frac{4n_{\rm ^{4}{\rm He}}}{n_{\rm n}+n_{\rm p}}\approx \frac{2n_{\rm n}}{n_{\rm n}+n_{\rm p}}$ (here $n_{\rm ^{4}{\rm He}}$, $n_{\rm n}$ and $n_{\rm p}$ are the number densities of the neutrons, protons and ${\rm ^{4}{\rm He}}$ respectively) is a very useful tool for studying the expansion rate of the Universe at the time of BBN, since it is very sensitive to temperature and hence to the expansion rate of the Universe at the time neutron-proton ratio freezes-out. In the \textit{standard BBN} (SBBN) for which it is assumed that the standard model of particle physics is valid (i.e., there are three families of neutrinos $N_{\nu}\approx 3$) and that the effective EoS of the physical content of the Universe during that time interval can be described by $p=\rho/3$, which gives the expansion rate of the Universe as $H_{\rm SBBN} = \frac{0.5}{t}$ through the Friedmann equations. We can utilize a good approximation for a primordial $^4$He mass fraction in the range $0.22\lesssim Y_{\rm p}\lesssim 0.27$ given by Steigman \cite{Steigman05,Steigman08} to predict $Y_{\rm p}$ values for non-standard expansion rates during BBN. Accordingly, if the assumption of the SBBN model expansion rate is relaxed, both BBN and CMB  will be affected and the approximation to $Y_{\rm p}$ in this case is given as follows:
\begin{equation}
\label{eqn:Steigman}
Y_{\rm p}=0.2485\pm 0.0006+0.0016[(\eta_{10}-6)+100\,(S-1)].
\end{equation}
where $S=H/H_{\rm SBBN}$ is the ratio of the expansion rate to the standard expansion rate and $\eta_{10}=10^{10}n_{\rm B}/n_{\gamma}$ is the ratio of baryons to photons in a comoving volume. We can safely ignore the term $\eta_{10}-6$ since observations give $\eta_{10}\sim 6$ and it is hundred times less effective than the term $S-1$. One may check that in the HEL model
\begin{equation}
H_{z\sim10^8}\cong H_{z\rightarrow\infty}\rightarrow\frac{\alpha}{t},
\end{equation}
{is a very good approximation. Hence, \eqref{eqn:Steigman} can safely be written as}
\begin{equation}
\label{eqn:Steigman_alpha}
Y_{\rm p}=0.2485\pm 0.0006+0.16\,(2\alpha-1),
\end{equation}
for the HEL model. Note that $S=2\alpha$ and SBBN is recovered provided that $S=2\alpha=1$. Using this equation with the value $\alpha=0.488_{-0.128}^{+0.124}$ ($1\sigma$) from Table \ref{table:HELobs}, that is obtained using the $H(z)$+SNe Ia data, we find that the predicted $^4$He abundance in the HEL model is
\begin{equation}
\label{eqn:qSteigman}
Y_{\rm p}=0.2448\pm 0.0450\quad (1\sigma).
\end{equation}

We note that this value covers both the SBBN value prediction $Y_{\rm p}^{\rm SBBN}=0.2485\pm 0.0006$ and also the most recent observational value $Y_{\rm p}=0.2534\pm 0.0083$ \cite{Aver12} that is obtained from the spectroscopic observations of the chemical abundances in metal-poor H II regions, an independent method for estimating the primordial helium abundance.

We should also examine if the age of the Universe was less than the lifetime of the free neutrons ($\tau_{\rm n}\sim887\,{\rm s}$) when the deuterium bottleneck would be broken, i.e., the CMB temperature drops down to $T\sim 80\,{\rm keV}$. Otherwise the BBN model would not work  properly and then our prediction given in \eqref{eqn:qSteigman} would not be valid. This can be done using the standard relation between the CBR temperature $T$ and the scale factor $a$ of the Universe in the HEL model:
\begin{equation}
\frac{a}{a_{0}}=\frac{T_{0}}{\eta T}=\left(\frac{t}{t_{0}}\right)^{\alpha}e^{\beta
\left(\frac{t}{t_{0}}-1\right)},
\end{equation}
where $\eta$ stands for any non-adiabatic expansion due to entropy production. In standard cosmology, the instantaneous $e^{\pm}$ annihilation is assumed at $T=m_{e}$. The heating due to this annihilation is accounted by $\eta$ where $\eta=1$ for $T<m_{e}$ while $\eta=(11/4)^{1/3}$ for $T>m_{e}$. It is enough for us to check whether the time scales are consistent and hence we simply consider $\eta=1$. Now using age of the present Universe in HEL model from Table \ref{table:HELLCDM1} and the present temperature of the Universe $T_{0}\cong 2.352\times 10^{-4}$eV ($T_{0}=2.728$ K) \cite{Fixsen09}, we find that the temperature $T\approx 80$ keV was reached when the age of the Universe was
\begin{equation}
t_{T=80\,{\rm keV}}=3.4969\pm39.3281\;{\rm seconds}.
\end{equation}
We note that this value is less than $\tau_{\rm n}\sim887\,{\rm seconds}$ and also very close to $\sim 1\,{\rm minute}$, the time scale that is expected for $T=80\,{\rm keV}$ in the conventional SBBN scenario.

It is interesting that using the cosmological data related with the expansion rate of the present day
Universe (spanning in the redshift range $0.015<z<1.750$), we predicted the dynamics of the early
Universe in the BBN epoch ($z\sim 10^{9}-10^{8}$) with a great success. This shows that although the HEL model
 fits the $H(z)$+SNe Ia data with a slightly less success compared to $\Lambda$CDM model,
 it has an advantage of describing the history of the Universe starting from the BBN epoch
 to the present day Universe whereas $\Lambda$CDM as well as many other dark energy models
 can describe the Universe only starting from the dust dominated epoch of the Universe.
 
 \subsection{BAO test}
\label{sec:today}

BAO observations provide a completely independent way from the supernova observations for investigating the expansion properties of the universe at low redshift values and give us the opportunity to compare and test the predictions of cosmological models at different redshift values. The imprint of the primordial baryon-photon acoustic oscillations in the matter power spectrum provides a \textit{standard ruler} via the dimensionless quantity \cite{Eisenstein05,Sollerman09}
\begin{equation}
A(z)=\sqrt{\Omega_{\rm m}}[H(z_1)/H_0]^{-1/3}\left[\frac{1}{z_1}\int_{0}^{z_1}\frac{H_{0}}{H(z)} {\rm d}z\right]^{2/3}.
\end{equation}
The BAO data set from the current surveys 6dFGS \cite{Beutler11}, SDSS \cite{Percival10}, and WiggleZ \cite{Blake11} spanning in the redshift range $0.106<z<0.73$, is shown in Table \ref{tab:BAO}, where we also give the predicted $A_{\rm HEL}(z)$ and $A_{\Lambda{\rm CDM}}(z)$ for the HEL and $\Lambda$CDM models respectively using the values for the parameters that we constrained using $H(z)$+SN Ia data (see Table \ref{table:HELobs}). We observe  that the predicted  $A_{\rm HEL}(z)$ and $A_{\Lambda{\rm CDM}}(z)$ values are consistent with each other as well as with the values from the BAO surveys.
\begin{table}[h]\centering\small
  \caption{The BAO data points from different surveys and their comparison with the $A(z)$ values predicted in HEL and $\Lambda$CDM models in our study.} 
    \begin{tabular}{llccc}
    \hline\hline
    $z$    &$A(z)$ from survey & $A_{\rm HEL}(z)$ & $A_{\Lambda{\rm CDM}}(z)$ \\
    \hline\hline
     0.106 & $0.526 \pm 0.028$ (6dFGS) & $ 0.517\pm0.001$ & $0.521\pm0.019$ \\[6pt]
    0.2    & $0.488 \pm 0.016$ (SDSS) & $ 0.499\pm0.002$ & $0.505\pm0.018$\\[6pt]
    0.35  & $0.484 \pm 0.016$ (SDSS)  & $ 0.472\pm0.003$ & $0.479\pm0.016$\\[6pt]
     0.44   &  $0.474 \pm 0.034$ (WiggleZ) & $ 0.457\pm0.003$ & $0.463\pm0.015$ \\[6pt]
     0.6    & $0.442 \pm 0.020$ (WiggleZ) & $ 0.431\pm0.003$ & $0.437\pm0.014$\\[6pt]
     0.73  & $0.424 \pm 0.021$ (WiggleZ) & $ 0.411\pm0.004$ & $0.416\pm0.013$ \\[6pt]
    \hline\hline \\
    \end{tabular}
  \label{tab:BAO}
\end{table}
 
\subsection{CMB test}
\label{sec:intermediate}

It is well known that the Universe should have transited from radiation- to matter-dominated era at $z\sim 3400$, and the recombination that leads to photon decoupling should have taken place at $z\sim 1100$ in the matter dominated era, i.e., where $w\simeq0$. This process should be achieved properly in a realistic description of the history of the Universe. A plot of the evolution of the effective EoS parameter in terms of redshift may be useful for a discussion in this respect. We plot the effective EoS parameters of HEL model (green curves) and $\Lambda$CDM model (red curves) versus redshift for $0<z<10^{4}$ in Figure \ref{figwz}. The solid green and red curves correspond to the mean values of the EoS parameters while the shaded regions between the dotted curves are $1\sigma$ error regions.
\begin{figure}[h]
\centering
{\psfrag{w}[b][b]{$w(z)$}
\psfrag{z}[b][b]{$z$}
\psfrag{m}[b][b]{\textcolor{red}{$w=\frac{1}{3}$}}
\psfrag{n}[b][b]{\textcolor{blue}{$w=-\frac{1}{3}$}}
\includegraphics[width=8.5 cm]{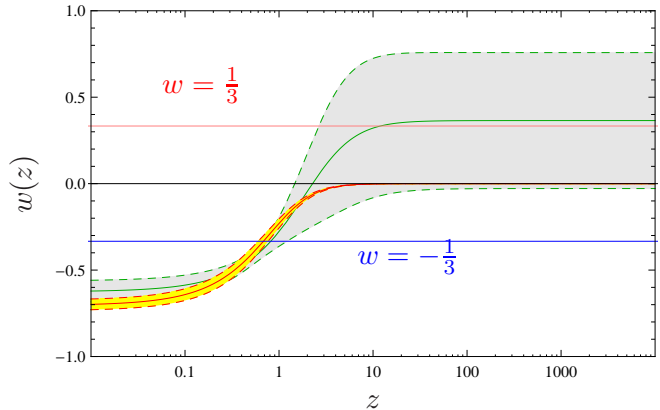} }
\caption{\footnotesize{The effective EoS parameters of HEL model (Green curves) and $\Lambda$CDM model (Red curves) are shown versus redshift with logarithmic scale on z-axis. The solid Green and Red curves correspond to the mean values of the EoS parameters while the shaded regions between the dotted curves are $1\sigma$ error regions. The Pink and Blue colored horizontal lines stand for $w=\frac{1}{3}$ and $w=-\frac{1}{3}$ respectively.}}
\label{figwz}
\end{figure}
We note first that there is, at high redshift values, a broad error region in HEL model but almost negligible error region in the $\Lambda$CDM model. In the later model, the error region shrinks as $z$ increases since the matter domination is the only possible past in this model. In the HEL model, on the other hand, the error region broadens as $z$ increases because error of the parameter $\alpha$ (determines essentially the early Universe) is larger than that of the parameter $\beta$ (determines essentially the late Universe) (see Table \ref{table:HELobs} and eqns. (\ref{eqn:early}) and (\ref{eqn:late})). We note that the mean value of the effective EoS parameter vanishes at $z\sim 2.5$ and remains almost firm at value $\sim \frac{1}{3}$ for the red-shift values higher than $z\sim 12$. This is obviously not in favor of the HEL model. On the other hand, within $1\sigma$ error region, it excludes neither a start with matter dominated era nor a start with radiation dominated era at $z\sim 1100$. However, this is because of the weaker constraints (larger error) on  the parameter $\alpha$ from $H(z)$+SN Ia data. We would evade this error completely by setting $\alpha=\frac{1}{2}$ so that $w\cong\frac{1}{3}$ at $z\sim 10^{8}-10^{9}$ in accordance with the SBBN. However, in this case, although the HEL model could describe the times of the BBN ($z\sim 10^{8}-10^{9}$) as well as the present times $z\sim 0$ successfully, it would not accommodate the matter dominated era properly, and hence would face complications with the CMB tests. 

A simple CMB test of the model may be done through the CMB shift parameter defined, in a spatially flat Universe, as
\begin{equation}
R=\sqrt{\Omega_{\rm m}}\int_{0}^{z_{dec}}\frac{H_{0}}{H(z)} {\rm d}z,
\end{equation}
where $\Omega_{\rm m}$ is the usual matter density parameter at the present time Universe, and the integral term is the comoving distance of the redshift $z_{\rm dec}$ at decoupling in a spatially flat Universe. This parameter describes the scaled distance to recombination, and is a useful tool for constraining and comparing models, which do not deviate lot from $\Lambda$CDM model \cite{Bond97,Elgaroy07,Sollerman09}. We adopt $z_{\rm dec}=1090$ for consistency with the latest observations (e.g., Planck experiment \cite{King13}) in our calculations. Using $\Omega_{\rm m}=0.29_{-0.02}^{+0.03}$ (the value obtained from the $H(z)+$SN Ia data in our study and consistent with Planck experiment), we find $R_{\Lambda{\rm CDM}}=1.73\pm0.07$ for the $\Lambda$CDM model. In HEL model, we do not have  explicit contribution from matter but we make use of the flat value $\Omega_{\rm m}=0.29$ since the $\Lambda$CDM and HEL models behave very close to each other at $z\sim0$ (see Figure \ref{figwz} and Table \ref{table:HELLCDM2}). Accordingly, using the values for the parameters $\alpha$ and $\beta$ from Table \ref{table:HELobs}, we find $R_{\rm HEL}=1.32\pm0.71$. The shift parameter we found for the $\Lambda$CDM model is consistent with the measured value $R_{\rm Planck} =1.744\pm 0.011$  in the Planck experiment \cite{Ade13,King13}. We note that, considering the mean values, the shift parameter predicted for the HEL model is way off these two values. On the other hand, it accommodates these two values within the error region, which is due to the large error in the parameter $\alpha$.

\bigskip
In the following section, we summarize the work done in this paper and then conclude it by discussing the possible directions for improving the HEL model in the light of the investigations done in the current section.

\section{Summary and future directions}
We have examined the hybrid form of scale factor, namely, a product of power law and an exponential function, which provides a  simple mechanism of transition from decelerating to accelerating phase. We showed that such an expansion history for the Universe can be obtained in the presence of dust for the particular case of Brans-Dicke theory of gravity. We also carried out the effective fluid and the single scalar field reconstruction using quintessence, tachyon and phantom fields, which can capture HEL in the framework of general relativity. We constrained the parameters of HEL model using the $25+580$ data points from the latest $H(z)$ and SN Ia compilations spanning in the redshift range $0.015<z<1.750$. We compared the kinematics and dynamics of the HEL model with that of the standard $\Lambda$CDM model. One may observe from the results displayed in Table \ref{table:HELLCDM1} that the two models are observationally indistinguishable in the vicinity of present epoch of the Universe. Statistically, we find $\Lambda$CDM model shows a slight better fit than the HEL model with the observational data. From the statefinder analysis of the HEL model in contrast with the other popular models such as Galileon, DGP, Chaplygin Gas and $\Lambda$CDM, we find that the HEL Universe evolves from radiation era to the de Sitter phase while the other models describe the Universe from the matter-dominated era to the de Sitter phase (see Fig. \ref{fig:rs}). The HEL model mimics the concordance $\Lambda$CDM behavior of the Universe at the present epoch. Using the values of model parameters obtained from the observational analysis, we extrapolated HEL beyond matter dominated era in the early Universe to the redshift values $z\sim 10^{9}-10^{8}$, where BBN proccesses are expected to occur. We find that the HEL model predicts the $^4$He abundance and time scale of the energy scales of the BBN processes with a great success. It is indeed interesting that the model is consistent with nucleo-synthesis which tells us that the simple expansion law under consideration can successfully describe thermal history as well as the late time transition to accelerating phase. The HEL model successfully passes the BAO test.

We conclude that the HEL and $\Lambda$CDM models are indistinguishable at low red-shift values. Also, the HEL model is good at describing the early radiation dominated era and the current accelerating phase of the Universe at the same time. However, from the CMB test we find that it does not accommodate the matter-dominated era properly unless we consider the parameter $\alpha$ with its large errors. Thus, with the current form of HEL, one should choose either to start the model with radiation domination that faces inconsistencies related with matter domination or start the model with matter domination at the expense of probing back to the radiation domination. The second case is not interesting since we know that $\Lambda$CDM model is doing pretty well. If one pursues the first option, one then needs to improve the model by modifying HEL ansatz such that a correction to it would cure the issues related with the intermediate matter-dominated era. In this respect two different routes may be followed: (a) One can use two power laws multiplying with an exponential term. This will bring additional free parameters, which is not fine as we have seen in AIC, KIC and BIC analysis in Section \ref{sec:obs}. However, one can choose, for instance, one of the powers such that the Universe will be dominated by radiation in the early times (minutes time scale). (b) One can use the potentials we obtained for a single scalar field driving the HEL expansion for describing the dark energy component in a cosmological model, where the well known components of the universe, such as matter and radiation, are given explicitly. In this case, the presence of matter and/or radiation in addition to the dark energy source described by the potential will give rise to a deviation from HEL ansatz. The state of our understanding of current acceleration of the Universe argues for keeping an open mind. Obtaining new forms of potential for describing dark energy source through the scale factor with some interesting properties like the HEL ansatz may provide us an opportunity to generate new classes of solutions that may fit the cosmological observations successfully.

%%%%%%%%%%%%%%%%%%%%%%%%%%%%%%%%%%%
%%%%%%%%%%%%%%%%%%%%%%%%%%%%%%%%%%%
%%%%%%%%%%%%%%%%%%%%%%%%%%%%%%%%%%%

% 

%%%%%%%%%%%%%%%%%%%%%%%%%%%%%%%%%%%
%%%%%%%%%%%%%%%%%%%%%%%%%%%%%%%%%%%
%%%%%%%%%%%%%%%%%%%%%%%%%%%%%%%%%%%
\bigskip

\begin{center}
\textbf{Acknowledgments}
\end{center}
We thank to J.D. Barrow, P. Diego and S.Y. Vernov for fruitful comments on the paper.
\"{O}.A. acknowledges the postdoctoral research scholarship he is receiving from The Scientific and Technological Research Council of Turkey (T\"{U}B{\.I}TAK-B{\.I}DEB 2218). \"{O}.A. appreciates also the support from Ko\c{c} University. S.K. is supported by the Department of Science and Technology, India under project No. SR/FTP/PS-102/2011. M.S. is supported by the Department of Science and Technology, India under project No. SR/S2/HEP-002/2008. L.X.'s work is supported in part by NSFC under the Grant No. 11275035 and ``the
Fundamental Research Funds for the Central Universities" under the Grant No. DUT13LK01. The authors are thankful to the anonymous referee for critical and fruitful comments on the paper.

\bigskip
\bigskip
\bigskip

%%%%%%%%%%%%%%%%%%%%%%%%%%%%%%%%%%%%%%%%%%%%%%%%%%%%%
\appendix

\section{The scalar field dynamics of $\Lambda$CDM model}
\label{sec:LCDM}

For the standard $\Lambda$CDM Universe, the scale factor, Hubble parameter, DP and jerk parameter read as
\begin{equation}
a=a_{1}\,\sinh^{\frac{2}{3}}\left(\sqrt{\frac{3\Lambda }{4}}\,t\right),
\quad
H=\sqrt{\frac{\Lambda }{3}}\coth\left(\sqrt{\frac{3\Lambda }{4}}\,t\right),
\quad
q=\frac{1}{2}-\frac{3}{2}\tanh^{2}\left(\sqrt{\frac{3\Lambda }{4}}\,t\right)
\quad\textnormal{and}\quad
j=1.
\end{equation}
The effective energy density and effective EoS parameter in the $\Lambda$CDM cosmology are given by
\begin{equation}
\rho_{\rm eff} = \Lambda\coth^2\left(\sqrt{\frac{3\Lambda }{4}}\,t\right)
\quad\textnormal{and}\quad
w_{\rm eff} = -\tanh^2\left(\sqrt{\frac{3\Lambda }{4}}\,t\right).
\end{equation}
The single canonical scalar field dynamics of $\Lambda$CDM model is described by
\begin{equation}
\phi(t)=\frac{2\sqrt{3}}{3}\ln\left[\tanh\left(\sqrt{\frac{3\Lambda }{16}}\,t\right)\right]+\phi_{ 1}
\quad\textnormal{and}\quad
V{(t)}=\frac{ \Lambda }{2 }\left[1+\text{coth}^2\left(\sqrt{\frac{3\Lambda }{4}}\,t\right)\right],
\end{equation}
where $\phi_{1}$ is the integration constant. The potential that generates such dynamics is
\begin{equation}
V(\phi)=\frac{ \Lambda }{4 }\left[3+\sinh\left\{\sqrt{3}(\phi-\phi_{1})\right\} \right].
\end{equation}


\begin{thebibliography}{99}
\bibitem{Riess98}
A.G. Riess et al. [Supernova Search Team Collaboration], Astron. J. \textbf{116}, 1009 (1998)   [arXiv:astro-ph/9805201].
\bibitem{Perlmutter99}
S. Perlmutter et al. [Supernova Cosmology Project Collaboration], Astrophys. J. \textbf{517}, 565 (1999) [arXiv:astro-ph/9812133].


%Dark energy reviews

%%%%%%
\bibitem{vpaddy}
P.J.E. Peebles and B. Ratra, Rev. Mod. Phys. \textbf{75}, 559 (2003) [astro-ph/0207347];
N. Straumann, gr-qc/0311083;
L. Perivolaropoulos, AIP Conf. Proc. \textbf{848}, 698 (2006) [astro-ph/0601014];
T. Padmanabhan, AIP Conf. Proc. \textbf{861}, 179 (2006) [astro-ph/0603114];
V. Sahni and A. Starobinsky, Int. J. Mod. Phys. D \textbf{15}, 2105 (2006) [astro-ph/0610026];
M. Sami, Lect. Notes Phys. \textbf{720}, 219 (2007);
M. Sami, arXiv:0901.0756 [hep-th];
J.A. Frieman, AIP Conf. Proc. \textbf{1057}, 87 (2008) [arXiv:0904.1832 [astro-ph.CO]];
S. Tsujikawa, arXiv:1004.1493 [astro-ph.CO].

\bibitem{review2}
E.J. Copeland, M. Sami and S. Tsujikawa, Int. J. Mod. Phys. D \textbf{15}, 1753 (2006) [hep-th/0603057].

\bibitem{review3}
E.V. Linder, Rept. Prog. Phys. \textbf{71}, 056901 (2008) [arXiv:0801.2968 [astro-ph]].

\bibitem{review3C}
R.R. Caldwell and M. Kamionkowski, Ann. Rev. Nucl. Part. Sci. \textbf{59}, 397 (2009) [arXiv:0903.0866 [astro-ph.CO]].

\bibitem{review3d}
A. Silvestri and M. Trodden, Rept. Prog. Phys. \textbf{72}, 096901 (2009) [arXiv:0904.0024 [astro-ph.CO]].

\bibitem{review4}
J. Frieman, M. Turner and D. Huterer, Ann. Rev. Astron. Astrophys. \textbf{46}, 385 (2008) [arXiv:0803.0982 [astro-ph]].

\bibitem{review5}
M. Sami, Curr. Sci. \textbf{97}, 887 (2009) [arXiv:0904.3445 [hep-th]].

\bibitem{Bamba:2012cp}
K. Bamba, S. Capozziello, S. Nojiri and S.D. Odintsov, Astrophys. Space Sci. \textbf{342}, 155 (2012) [arXiv:1205.3421 [gr-qc]].


%Modified gravitiy reviews
\bibitem{Nojiri11}
S. Nojiri and S.D. Odintsov, Phys. Rep. \textbf{505}, 59 (2011)     [arXiv:1011.0544 [gr-qc]].
\bibitem{Clifton12}
T. Clifton, P.G. Ferreira, A. Padilla and C. Skordis, Phys. Rep.
\textbf{513}, 1 (2012) [arXiv:1106.2476 [astro-ph.CO]].
\bibitem{sf1}T. Padmanabhan, Phys. Rev. D \textbf{66}, 021301 (2002)
[arXiv:hep-th/0204150].
\bibitem{sf2}T. Padmanabhan and T.R. Choudhury, Phys. Rev. D \textbf{66} 081301 (2002) [arXiv:hep-th/0205055].
\bibitem{sf3}J.S. Bagla, H.K. Jassal and T. Padmanabhan, Phys. Rev. D \textbf{67}, 063504 (2003)
[arXiv:astro-ph/0212198].
\bibitem{sf4} A.Yu. Kamenshchik, A. Tronconi, G. Venturi, S.Yu.
Vernov, Phys. Rev. D \textbf{87}, 063503 (2013) [arXiv:1211.6272].
\bibitem{sf5} S.V. Chervon, V.M. Zhuravlev and V.K. Shchigolev, Phys. Lett. B \textbf{398}, 269 (1997)
[arXiv:gr-qc/9706031].
\bibitem{sf6} M. Sami, P. Chingangbam and T. Qureshi, Pramana \textbf{62}, 765 (2004)
[arXiv:hep-th/0301140].
\bibitem{sf7}M. Sami, Mod. Phys. Lett. A \textbf{18}, 691 (2003) [arXiv:hep-th/0205146].
\bibitem{sf8}S.V. Chervon and V.M. Zhuravlev, arXiv:gr-qc/9907051.
\bibitem{staro}T.D. Saini, S. Raychaudhury , V. Sahni and A.A.
Starobinsky, Phys.Rev.Lett. \textbf{85}, 1162 (2000) [arXiv:astro-ph/9910231].
\bibitem{polar} M. Chevallier and D. Polarski, Int. J. Mod. Phys. D \textbf{10}, 213 (2001) [arXiv:gr-qc/0009008].
\bibitem{linder} E.V. Linder,   Phys. Rev. Lett. \textbf{90}, 091301 (2003) [arXiv:astro-ph/0208512].
\bibitem{wparam}
Y. Gong and A. Wang, Phys. Rev. D \textbf{75}, 043520 (2007) [arXiv:astro-ph/0612196];
Y. Gong, X. Zhu and Z. Zhu, MNRAS \textbf{415}, 1943 (2011) [arXiv:1008.5010];
R. Erdem, arXiv:1105.0345 [gr-qc];
H. Li and X. Zhang, JCAP \textbf{1205}, 029 (2012) [arXiv:1106.5658];
A. De Felice, S. Nesseris and S. Tsujikawa, [arXiv:1203.6760];
S. Campo, I. Duran, R. Herrera and D. Pavon,  Phys. Rev. D \textbf{86}, 083509 (2012) [arXiv:1209.3415];
E. Elizalde, E.O. Pozdeeva and S.Y. Vernov, Class. Quantum Grav. \textbf{30} 035002 (2013)  [arXiv:1209.5957 [astro-ph.CO]];
A. Aviles, C. Gruber, O. Luongo and H. Quevedo, arXiv:1301.4044;
E.R.M. Tarrant, E.J. Copeland, A. Padilla and C. Skordis, arXiv:1304.5532.

%%%%%%%
%Stiff fluid for the very very early Universe
\bibitem{Zeldovich62}
Y.B. Zeldovich, Soviet Physics JETP \textbf{14}, 1143 (1962).
\bibitem{Barrow78}
J.D. Barrow, Nature \textbf{272}, 211 (1978).

\bibitem{Duff95}
M.J. Duff, R.R. Khuri, J.X. Lu, Phys. Rept. \textbf{259}, 213-326 (1995) [arXiv:hep-th/9412184].
\bibitem{Lidsey00}
J.E. Lidsey, D. Wands, E.J. Copeland, Phys. Rept. \textbf{337} 343-492 (2000) [arXiv:hep-th/9909061].

\bibitem{Parsons}
P. Parsons and J.D. Barrow, Class. Quantum Grav. \textbf{12} 1715 (1995).

\bibitem{Gumjudpai12}
B. Gumjudpai and C. Kaeonikhom, arXiv:1201.3499.

\bibitem{Picon00}
C. Armendariz-Picon, V.F. Mukhanov and P.J. Steinhardt, Phys. Rev. Lett. \textbf{85}, 4438 (2000) [arXiv:astro-ph/0004134].
\bibitem{Gibbons02}
G.W. Gibbons, Phys. Lett. B \textbf{537}, 1 (2002) [arXiv:hep-th/0204008].
\bibitem{Chimento04}
L.P. Chimento, Phys. Rev. D \textbf{69}, 123517 (2004) [arXiv:astro-ph/0311613].
\bibitem{Bagla03}
J.S. Bagla, H.K.Jassal and T. Padmanabhan, Phys. Rev. D \textbf{67}, 063504 (2003) [arXiv:astro-ph/0212198].
\bibitem{Copeland05}
E.J. Copeland, M.R. Garousi, M. Sami and S. Tsujikawa, Phys. Rev. D \textbf{71}, 043003 (2005) [arXiv:hep-th/0411192].
\bibitem{Calcagni06}
G. Calcagni and A.R. Liddle, Phys. Rev. D \textbf{74}, 043528 (2006) [arXiv:astro-ph/0606003].

\bibitem{Vazquez12}
J.A. Vazquez, M. Bridges, M.P. Hobson and A.N. Lasenby, JCAP \textbf{09}, 020 (2012) [arXiv:1205.0847 [astro-ph.CO]].
\bibitem{Nobosyadlyj12}
B. Novosyadlyj, O. Sergijenko, R. Durrer and V. Pelykh, Phys. Rev. D \textbf{86}, 083008 (2012) [arXiv:1206.5194 [astro-ph.CO]].
\bibitem{Parkinson12}
D. Parkinson et al., Phys. Rev. D \textbf{86}, 103518 (2012) [arXiv:1210.2130 [astro-ph.CO]].
\bibitem{Ade13}
P.A.R. Ade et al. [Planck Collaboration], arXiv:1303.5076 [astro-ph.CO]


\bibitem{Shtanov03}
V. Sahni and Y. Shtanov, JCAP \textbf{11}, 014 (2003) [arXiv:astro-ph/0202346].
\bibitem{Elizade04}
E. Elizalde, S. Nojiri and S.D. Odintsov, Phys. Rev. D \textbf{70}, 043539 (2004) [arXiv:hep-th/0405034].

\bibitem{Chen02}
C.M. Chen, D.V. Gal’tsov and M. Gutperle, Phys. Rev. D \textbf{66}, 024043 (2002) [arXiv:hep-th/0204071].
\bibitem{Townsent03}
P.K. Townsend, M.N.R. Wohlfarth, Phys. Rev. Lett. \textbf{91}, 061302 (2003) [arXiv:hep-th/0303097].
\bibitem{Ohta03}
N. Ohta, Phys. Rev. Lett. \textbf{91}, 061303 (2003) [arXiv:hep-th/0303238].

\bibitem{Caldwell02}
R.R. Caldwell, Phys. Lett. B \textbf{545}, 23 (2002) [arXiv:astro-ph/9908168].

%Data


\bibitem{ref:SVJ2005}
J. Simon, L. Verde and R. Jimenez, \textit{Phys. Rev. D} \textbf{71} (2005) 123001 [arXiv:astro-ph/0412269].

\bibitem{Gaztanaga09}
E. Gaztanaga, A. Cabre and L. Hui, \textit{Mon. Not. R. Astron. Soc.} \textbf{399} (2009) 1663 [arXiv:0807.3551 [astro-ph]].

\bibitem{stern10}
D. Stern, R. Jimenez, L. Verde, M. Kamionkowski and S. A. Standford, \textit{J. Cosmol. Astropart. Phys.} \textbf{02} (2010) 008 [arXiv:0907.3149 [astro-ph.CO]].
\bibitem{moresco12}
M. Moresco et al., \textit{J. Cosmol. Astropart. Phys.} \textbf{08} (2012) 006 [arXiv:1201.3609 [astro-ph.CO]].

\bibitem{ref:Zhang2012} C. Zhang, et al., arXiv:1207.4541 [astro-ph.CO].


\bibitem{lvdp13}
O. Akarsu, T. Dereli, S. Kumar and L. Xu,  arXiv:1305.5190.

\bibitem{ref:Suzuki2012} N. Suzuki, et al.,  Astropart. Phys. \textbf{746}, 85 (2012) [arXiv:1105.3470 [astro-ph.CO]].
\bibitem{ref:MCMC} A. Lewis and S. Bridle, Phys. Rev. D \textbf{66}, 103511 (2002). (http://cosmologist.info/cosmomc/)


%\bibitem{Li2012}
%Z. Li, P. Wu and H. Yu, \textit{ApJ} \textbf{744} (2012) 176 [arXiv:1109.6125 [astro-ph.CO]].

\bibitem{Melia2013}
F. Melia and R. S. Maier,  	arXiv:1304.1802 [astro-ph.CO]
%JERK
\bibitem{sahni03}
V. Sahni,  T.D. Saini, A.A. Starobinsky and U. Alam, JETP Letters \textbf{77}, 201 (2003) [arXiv:astro-ph/0201498].
\bibitem{Visser04}
M. Visser,  Classical Quant. Grav. \textbf{21}, 2603 (2004) [arXiv:gr-qc/0309109].
\bibitem{Cattoen08}
C. Cattoen and M. Visser,  Phys. Rev. D \textbf{78}, 063501 (2008) [arXiv:0809.0537 [gr-qc]].
\bibitem{Wang09}
F.Y. Wang, Z.G. Dai and S. Qi,  Astron. Astrophys. \textbf{507}, 53 (2009) [arXiv:0912.5141 [astro-ph.CO]].
\bibitem{Vitagliano10}
V. Vincenzo et al.,  JCAP \textbf{03}, 005 (2010).
\bibitem{Capozziello11}
S. Capozziello, R. Lazkoz and V. Salzano,  Phys. Rev. D \textbf{84}, 124061 (2011) [arXiv:1104.3096 [astro-ph.CO]].
\bibitem{Xia12}
J.Q. Xia, V. Vitagliano, S. Liberati and M. Viel, Phys. Rev. D \textbf{85},  043520 (2012) [arXiv:1103.0378 [astro-ph.CO]].

\bibitem{Sami12}
M. Sami, M. Shahalam, M. Skugoreva and A. Toporensky,  Phys. Rev. D \textbf{86}, 103532 (2012) [arXiv:1207.6691 [hep-th]].

% Early Universe
\bibitem{Steigman05}
G. Steigman, Int. J. Mod. Phys. E \textbf{15}, 1 (2006) [arXiv:astro-ph/0511534].
\bibitem{Steigman08}
V. Simha and G. Steigman,  JCAP \textbf{06}, 016 (2008) [arXiv:0803.3465 [astro-ph]].
\bibitem{Aver12}
E. Aver, K.A. Olive and E.D. Skillman,  JCAP \textbf{04}, 004 (2012) [arXiv:1112.3713 [astro-ph.CO]].
\bibitem{Fixsen09}
D. J. Fixsen,  Astrophys. J. \textbf{707}, 916 (2009) [arXiv:0911.1955 [astro-ph.CO]].

%BAO
\bibitem{Eisenstein05}
D.J. Eisenstein, \textit{Astrophys. J.} \textbf{633}, 560 (2005) [arXiv:astro-ph/0501171]
\bibitem{Sollerman09}
J. Sollerman et al., \textit{Astrophys. J.} \textbf{703} 1374 (2009) [arXiv:0908.4276 [astro-ph.CO]]
\bibitem{Beutler11}
F. Beutler et al., \textit{Mon. Not. R. Astron. Soc.} \textbf{416}, 3017 (2011) [arXiv:1106.3366 [astro-ph.CO]].
\bibitem{Percival10}
W.J. Percival et al., \textit{Mon. Not. R. Astron. Soc.} \textbf{401},  2148 (2010) [arXiv:0907.1660 [astro-ph.CO]]
\bibitem{Blake11}
C. Blake et al., \textit{Mon. Not. R. Astron. Soc.} \textbf{418}, 1707 (2011) [arXiv:1108.2635 [astro-ph.CO]].

%Shift parameter
\bibitem{Bond97}
J.R. Bond, G. Efstathiou, M. Tegmark, \textit{Mon. Not. R. Astron. Soc.} \textbf{291}, L33 (1997) [arXiv:astro-ph/9702100].
\bibitem{Elgaroy07}
{\O}. Elgar{\o}y, T. Multam\"{a}ki, \textit{Astron. Astrophys.} \textbf{471} 65 (2007) [arXiv:astro-ph/0702343]
\bibitem{King13}
A.L. King et al., arXiv:1311.2356 [astro-ph.CO]


\end{thebibliography}
\end{document}